\documentclass[11pt,a4paper]{article}
\usepackage{graphicx}  % needed for figures
\usepackage{dcolumn}   % needed for some tables
\usepackage{bm}        % for math
\usepackage{amssymb}   % for math
\usepackage[latin1]{inputenc}
\usepackage{amsmath}
\usepackage{amsfonts}
\usepackage{amssymb}
\usepackage{setspace}
\usepackage{mathtools}
\usepackage{pstricks}
\usepackage{color}
\usepackage{bbold}
\usepackage{jcappub}
\providecommand{\f}[2]{\frac{{#1}}{{#2}}}
\newcommand{\da}{\ensuremath{\dot{a}}}
\newcommand{\dda}{\ensuremath{\ddot{a}}}

\newcommand{\ee}[1]{\begin{equation}#1\end{equation}}
\newcommand{\ea}[1]{\begin{align}#1\end{align}}

\def\undertilde#1{\mathord{\vtop{\ialign{##\crcr
$\hfil\displaystyle{#1}\hfil$\crcr\noalign{\kern1.5pt\nointerlineskip}
$\hfil\tilde{}\hfil$\crcr\noalign{\kern1.5pt}}}}}

\title{Decoherence Can Relax Cosmic Acceleration}

\author[a]{Tommi Markkanen}
\affiliation[a]{Department of Physics, King's College London, Strand, London WC2R 2LS, UK}

\abstract{In this work we investigate the semi-classical backreaction for a quantised conformal scalar field and classical vacuum energy. In contrast to the usual approximation of a closed system, our analysis includes an environmental sector such that a quantum-to-classical transition can take place. We show that when the system decoheres into a mixed state with particle number as the classical observable de Sitter space is destabilized, which is observable as a gradually decreasing Hubble rate. In particular we show that at late times this mechanism can drive the curvature of the Universe to zero and has an interpretation as the decay of the vacuum energy demonstrating that quantum effects can be relevant for the fate of the Universe.}

\emailAdd{tommi.markkanen@kcl.ac.uk}

\begin{document}

\maketitle
%- references
\section{Introduction}
Devising a quantum theory of gravity has turned out to be an extremely difficult problem, which to this day remains without a complete solution. As originally suggested in \cite{mol,ros}, a useful approximation for the study of the backreaction of the quantum matter onto the spacetime geometry is obtained by treating space-time classically and using the expectation value of the quantised energy-momentum tensor as the source term in the Einstein equation 
\ee{G_{\mu\nu}= \f{8\pi G}{c^4}\langle\hat{T}_{\mu\nu}\rangle\,.\label{eq:semiE}}
This semi-classical approach has many applications \cite{Birrell:1982ix}, most notably it may be used to predict black hole evaporation \cite{Hawking:1974sw}. In conditions where quantum gravity is unimportant this framework is expected to give reliable results, although it is not without issues \cite{Kuo:1993if,Kibble:1980ia,eppley}. 

Usually to a good approximation systems of macroscopic size can be described in terms of classical physics, despite the fact that the classical configurations that are frequently encountered in Nature are only a small subset of all possible quantum states: quantum interference, or more descriptively, "cat states" named after Schr\"{o}dinger's famous thought experiment are generally absent.

In the decoherence program the emergence of a classical reality from the quantum realm i.e. the quantum-to-classical transition is explained by using no other mechanisms than what are provided by quantum theory \cite{Zeh:1970zz,Zurek:1982ii,Zurek:1981xq,Joos:1984uk}, see \cite{Schlosshauer:2003zy,Zurek:2003zz,Zeh:1995jg,Kiefer:1997hv,Zurek:1991:vv} for a historical account and a more complete list of references. The key element is the realization that a macroscopic quantum system is never completely isolated or closed and inevitably its wave function will be influenced by environmental degrees of freedom, which themselves are not directly observable. This results in a change of the systems characteristics,
where the quantum phases decohere leading to a suppression of the interference terms while the eigenstates of the classical observables remain intact. In a way decoherence sifts through the Hilbert space of all states filtering out the ones that are the most stable under environmental monitoring \cite{Kuebler:1973mm,Zurek:1992mv,Zurek:1993:pu,Zurek:1998ji}.
Hence, decoherence provides an explanation why classicalization occurs and which particular observables are the robust classical ones. In the inflationary paradigm, decoherence is vital for the predictions for the large scale structure of the Universe in that it provides a mechanism by which initially quantum fluctuations transform into classical perturbations \cite{Polarski:1995jg,Kiefer:1998qe,Kiefer:2006je,Campo:2008ju,Campo:2008ij,
Kiefer:2008ku,Burgess:2006jn,Burgess:2014eoa,Martineau:2006ki,Nelson:2016kjm}. This highlights a more general crucial feature of quantum theory: in order to obtain predictions that are in accord with the classical world around us, one must often consider an open quantum system, see for example \cite{Calzetta:2008iqa,Giulini:1996nw} for more discussion. As one of the main points of this work we argue that an unobservable environmental sector and in particular decoherence can also qualitatively change the predictions of semi-classical gravity, as advocated also in \cite{Tegmark:2011pi}.

Some of the issues of closed quantum systems can be understood by considering the evolution of pure states: unitarity forces a pure state to always remain pure and hence to contain no entropy. For the many examples of cosmological particle creation \cite{Bernard:1977pq,Parker,Audretsch:1978qu, Parker:1968mv,Parker:1968mv1,Parker:1968mv3} this implies that the system contains correlations not present in a thermal distribution, despite a thermal average particle number \cite{Kandrup:1988sg,Kandrup:1988zv, Hu:1986jd,Parker:2009uva,Hu:1986jj}. For example, in the Unruh effect \cite{Unruh:1976db} where an accelerated observer sees the Minkowski space vacuum -- a pure state -- as thermally occupied, a truly thermal state is obtained only after a coarse graining over the unobservable environment is performed \cite{Crispino:2007eb,Unruh:1983ms}. A similar step is also required for obtaining a thermal density matrix for black holes \cite{Hawking:1976ra}.
An increase in entropy can be seen as loss of information \cite{Shannon:1948zz,Jaynes:1957zza} and is a generic result of coarse graining because an open system can disperse information from the system into the environment. Open quantum systems in particle creation and semi-classical backreaction are investigated in \cite{Calzetta:1993qe,Hu:2008rga,Calzetta:1990zu, Calzetta:1999zr,Paz:1991nd,Anderson:2005hi, Calzetta:1995ys, Kiefer:2001wn} and coarse graining in the cosmological context  is studied in \cite{Kandrup:1988zv,Hu:1986jd,Hu:1986jj,Brandenberger:1992jh,Prokopec:1992ia, Gasperini:1993yf,KeskiVakkuri:1993vk,Lin:2010pfa}.

Due to the highly symmetric nature of de Sitter space, its stability in the quantised theory has been the subject of considerable interest \cite{Tsamis:1996qq2,Padmanabhan:2002ha,Tsamis:1996qq3,Geshnizjani:2002wp,Unruh:1998ic,Geshnizjani:2003cn,Tsamis:1996qq, Tsamis:1996qq0,Tsamis:1996qq1,Brandenberger:1999su,ford,Bousso:2001mw,Torrieri:2015kla,Polyakov:2007mm,Polyakov:2007mm1, Polyakov:2007mm2,akh,Marolf:2010zp,Marolf:2010nz,Marozzi:2006ky,Marozzi:2012tp,Finelli:2011cw,Marozzi:2014xma,ArkaniHamed:2007ky,Greene:2005wk, Rigopoulos:2016oko,Firouzjaee:2015bqa,clif,Antoniadis:1985pj,Markkanen:2016vrp,Mottola:1984ar,Mottola:1985qt, Anderson:2013ila,Anderson:2013ila1,Kachru:2003aw,Goheer:2002vf,Sekiwa:2006qj,Padmanabhan:2003gd,Koivisto:2010pj, Albrecht:2014hxa,Rajaraman:2016nvv}. In the semi-classical approach the global de Sitter manifold that includes a contracting phase has been argued to be unstable in \cite{Mottola:1984ar,Anderson:2013ila,Anderson:2013ila1}. However, in the current cosmological framework the primary interest is in the exponentially expanding patch of de Sitter space as given by the flat Friedmann--Lema\^{i}tre--Robertson--Walker (FLRW) %\footnote{Friedmann--Lema\^{i}tre--Robertson--Walker} 
coordinates. In the expanding patch it has been shown that at least in the semi-classical approach for a closed system with a non-interacting scalar field and a non-tachyonic effective mass the system equilibrates to a de Sitter invariant state, the Bunch-Davies vacuum, and no instabilities arise \cite{Anderson:2000wx,markrajan,Habib:1999cs,Albrecht:2014hxa}. But in a scenario where the Bunch-Davies state is allowed to decohere it is not obvious that de Sitter symmetry can be respected. In fact, a violation of de Sitter symmetry from decoherence has been noted to arise when the inflationary perturbations classicalise \cite{Kiefer:2008ku}.

Via the semi-classical approach we will study the effects from decoherence in a model consisting of a conformal quantum scalar field and classical vacuum energy, with a focus on the implications for the fate of a Universe dominated by dark energy. When the system is closed this scenario leads to eternal exponential expansion, but we argue that decoherence can change this conclusion.

We will work in the units $\hbar\equiv c\equiv k_B\equiv1$ and with the conventions (+,+,+) of \cite{Misner:1974qy}.
\section{Decoherence, entropy and coarse graining}
\label{sec:dec}
In this section we give the essential features of decoherence which we need in the calculation of section \ref{sec:sta}. For more details we refer the reader to the reviews and textbooks \cite{Giulini:1996nw,Zurek:1991:vv,Zeh,Schlosshauer:2007:un,Schlosshauer:2003zy,Zurek:2003zz,Kiefer:1997hv,Zeh:1995jg}.

Let us first suppose the existence of a quantum system to which we have experimental access initialized in the pure state $\vert \Psi\rangle$. In some basis defined by $\vert n\rangle$, which we assume to be orthonormal, we can then write the initial density matrix $\hat{\rho}$ as
\ee{\hat{\rho}=\vert\Psi\rangle\langle\Psi\vert=\sum_n\sum_m c(n)c^*(m)\vert n\rangle\langle m\vert\,;\qquad c(n)\equiv\langle n \vert \Psi\rangle
\,,}
with the normalization
\ee{\qquad{\rm Tr}\big\{\hat{\rho}\big\}=1\quad\Leftrightarrow\quad \sum_n \vert c(n) \vert^2 =1\,.}
An open system often exhibits suppression of the off-diagonal terms, which in the $\vert n \rangle$ basis we can schematically write as
\ee{\hat{\rho}\quad%\overset{decoherence}{\longrightarrow}
\longrightarrow\quad\hat{\rho}^\text{\tiny D}=\sum_n \vert c(n)\vert^2\vert n\rangle\langle n\vert\,,\label{eq:dec0}}
where the superscript "D" stands for \textit{decohered}. The dissappearance of the off-diagonal or quantum interference terms in the density matrix (\ref{eq:dec0}) is the defining characteristic of decoherence. The basis $\vert n\rangle$ is  chosen by the specific form of environmental interaction and is referred to as the pointer basis \cite{Zurek:1981xq}.

As we can see from (\ref{eq:dec0}), the decohered system is mathematically equivalent to a classical statistical ensemble for which we can define entropy. If we initially have a pure state with $\hat{\rho}^2=\hat{\rho}$, which then decoheres into a mixed state with \ee{(\hat{\rho}^\text{\tiny D})^2\neq\hat{\rho}^\text{\tiny D}\quad\Leftrightarrow\quad0\leq\vert c(n) \vert^2<1\,,}
the von Neumann entropy will increase in the process as
\ee{S=-{\rm Tr}\big\{\hat{\rho}\log\hat{\rho}\big\}=0\quad%\overset{decoherence}{\longrightarrow}
\longrightarrow\quad-{\rm Tr}\big\{\hat{\rho}^{\text{\tiny D}}\log\hat{\rho}^{\text{\tiny D}}\big\}=-\sum_n \vert c(n)\vert^2\log \vert c(n)\vert^2>0\,.\label{eq:en}}
In a pure state one has the maximum allowed information of the quantum system, which is related to the fact that a pure state has no entropy. An increase in entropy means loss of information and implies that $\hat{\rho}^{\text{\tiny D}}$ is a coarse grained version of the original $\hat{\rho}$, which can seem problematic when considering the unitary evolution of quantum theory. This is resolved by the realization that $\hat{\rho}^{\text{\tiny D}}$ is only a part of the global state, which includes also the environment and for which there is no violation of unitarity. The loss of information may then be understood to arise via entanglement where the systems quantum phase relations are delocalized into correlations between the system and the environment, which are unobservable. To be precise, in our notation $\hat{\rho}^{\text{\tiny D}}$ denotes a reduced density matrix derived from the global one by tracing over the environment.

In this work decoherence is defined as the disappearance of the off-diagonal terms of the density matrix, irrespective of the cause. One may broadly categorize such an evolution to be the result of two distinct scenarios: (1) in addition to the system there are unobservable degrees of freedom forming an environment which can act as a sink for information, (2) a local observer loses observational access to some of the degrees of freedom in the initial system, which then may effectively be viewed as the environment.

Next we need an explicit expression for the decohered density matrix $\hat{\rho}^\text{\tiny D}$ in the quantum field theory setting and for that we need to know the pointer basis. In principle, it is possible to determine the pointer basis from first principles by defining a system, an environment and their interaction and explicitly studying their dynamics \cite{Joos:1984uk,Caldeira:1982iu,Hu:1991di}. Another popular method is to implement the "predictability sieve" technique \cite{Zurek:1992mv,Zurek:1993:pu}. Our approach will be to study the implications of a particular physically motivated choice for $\hat{\rho}^\text{\tiny D}$.

For a free quantum field a natural representation is given by the Fock space consisting of an infinite number of harmonic oscillators, which also provides a set of observables with an obvious connection to classical physics, the number of quanta per given co-moving momentum $n_\mathbf{k}$. Our experiences as observers inhabiting an expanding space indicate that particle number in a weakly curved space is to a good approximation a well-defined classical observable such that macroscopic objects are found in states with definite $n_\mathbf{k}$. Simply put, we know that in the classical world we are able to count particles. Based on this notion, the assumption we will make is that interference between states with different particle number is suppressed in our system. This will result in a density matrix with an interpretation as a statistical ensemble of states with definite particle number\footnote{In curved space there is an important subtlety here, as the definition of a particle is inherently ambiguous. This matter is addressed in section \ref{sec:sta}}. This density matrix can then symbolically be expressed as
\ee{%\hat{\rho}\quad\overset{decoherence}{\longrightarrow} \quad
\hat{\rho}^{\text{\tiny D}}=\sum_{n_{\mathbf{k}_1}=0}^\infty\sum_{n_{\mathbf{k}_2}=0}^\infty\cdots\vert \mathcal{C}(n_{\mathbf{k}_1},n_{\mathbf{k}_2},\cdots)\vert^2\vert n_{\mathbf{k}_1},n_{\mathbf{k}_2},\cdots\rangle\langle\cdots, n_{\mathbf{k}_2},n_{\mathbf{k}_1}\vert\, ,\label{eq:decQFT0}} 
where the sequence $(n_{\mathbf{k}_1},n_{\mathbf{k}_2},\cdots)$ is to be understood to span all momenta and the $\mathcal{C}$'s are constrained by the normalization requirement ${\rm Tr}\big\{\hat{\rho}^{\text{\tiny D}}\big\}=1$. Note that the zero mode $\mathbf{k}=0$ is not included in (\ref{eq:decQFT0}) as for a conformal scalar field these states are indistinguishable from the vacuum, much like for a photon. 
This specific coarse graining where only states with definite particle number appear has previously been studied in various contexts in \cite{Hu:1986jd,Prokopec:1992ia,Hu:1986jj,Kandrup:1988sg,Kandrup:1988zv,Lin:2010pfa} and is sometimes called the random phase approximation. 

The classicalization of the inflationary perturbations via decoherence is usually assumed to take place in the field amplitude basis \cite{Polarski:1995jg,Kiefer:2006je,Kiefer:2008ku,Burgess:2006jn}, and not in the particle number basis as described by (\ref{eq:decQFT0}). This is due to the fact that the scalar field responsible for the generation of the primordial perturbations is not conformally invariant and because of this its modes experience significant squeezing when crossing the horizon. However, we only consider the case of a conformally coupled field for which there is no squeezing. We also point out that in the important special case where information loss is due to the existence of a horizon, the result after coarse graining is precisely a density matrix diagonal in the particle number basis \cite{Hawking:1976ra,Unruh:1983ms,Markkanen:2016vrp}.

An additional physically motivated coarse graining can be introduced by assuming further that the different occupation numbers are not correlated, so that the decohered density matrix of the system can be constructed as a product of the density matrices of the individual oscillators. This can be obtained from (\ref{eq:decQFT0}) by
\ee{%\hat{\rho}\quad\overset{decoherence}{\longrightarrow} \quad
\vert \mathcal{C}(n_{\mathbf{k}_1},n_{\mathbf{k}_2},\cdots)\vert^2=\prod_\mathbf{k}\vert c(n_\mathbf{k},{\mathbf{k}})\vert^2\quad\Rightarrow\quad\hat{\rho}^{\text{\tiny D}}=\bigotimes_{\mathbf{k}}\sum_{n_{\mathbf{k}}=0}^\infty\vert c(n_\mathbf{k},{\mathbf{k}})\vert^2\vert n_{\mathbf{k}}\rangle\langle n_{\mathbf{k}}\vert\equiv\bigotimes_{\mathbf{k}}\hat{\rho}^{\text{\tiny D}}_\mathbf{k}\, ,\label{eq:decQFT}}
%\ee{\hat{\rho}=\bigotimes_{\mathbf{k}}\hat{\rho}_\mathbf{k}\,\qquad ;\,\hat{\rho}_\mathbf{k}= \sum_{n_{\mathbf{k}}=0}^\infty\sum_{m_{\mathbf{k}}=0}^\infty c(n_\mathbf{k},{\mathbf{k}})c^*(m_\mathbf{k},{\mathbf{k}})\vert n_{\mathbf{k}}\rangle\langle m_{\mathbf{k}}\vert\,,\label{eq:dec1}}
%each oscillator is expanded in terms of  particle number eingenstates $\vert n_{\mathbf{k}}\rangle$ given by some yet undefined mode\footnote{In (\ref{eq:dec1}) we need the additional $\mathbf{k}$ dependence as we want each oscillator to potentially have a different density matrix.}. 
where $\hat{\rho}^{\text{\tiny D}}_\mathbf{k}$ is the decohered density matrix for a single oscillator. Here the normalization constraint gives
\ee{{\rm Tr} \big\{\hat{\rho}^{\text{\tiny D}}\big\}=1\quad\Leftrightarrow\quad {\rm Tr} \big\{\hat{\rho}^{\text{\tiny D}}_\mathbf{k}\big\}=1\quad\Leftrightarrow\quad\sum_{n_{\mathbf{k}}=0}^\infty \vert c(n_\mathbf{k},{\mathbf{k}})\vert^2 =1\,.\label{eq:decchoi}}
By using (\ref{eq:decQFT}) we can show a desirable property possessed by (\ref{eq:decQFT0}), which is that it includes a thermal spectrum as a special case, $\vert c(n_\mathbf{k},{\mathbf{k}})\vert^2=Z^{-1}e^{- n_\mathbf{k}\vert\mathbf{k}\vert/T}$ with $Z$ given by the normalization  (\ref{eq:decchoi}). For a thermal distribution the standard black-body result for the entropy of a scalar field $S=(2\pi^2/45)VT^3$ easily follows from (\ref{eq:en})\footnote{It is also straightforward to show that a mixture of coherent states with a randomized phase as
\ee{\hat{\rho}_\mathbf{k}^\text{\tiny D}=\f{1}{2\pi}\int^{2\pi}_0 d\theta_\mathbf{k}\,\big\vert|\alpha_\mathbf{k}|e^{i\theta_\mathbf{k}}\big\rangle\big\langle e^{i\theta_\mathbf{k}}|\alpha_\mathbf{k}|\big\vert\,;\qquad\big\vert\alpha_\mathbf{k}\big\rangle=e^{-|\alpha_\mathbf{k}|^2/2}\sum_{n_\mathbf{k}=0}^\infty\f{\alpha_\mathbf{k}^{n_\mathbf{k}}}{\sqrt{n_\mathbf{k}!}}\big\vert n_\mathbf{k}\big\rangle,}
is contained in (\ref{eq:decQFT}) and hence in (\ref{eq:decQFT0}).}.

For the coarse graining (\ref{eq:decQFT0}) one has the natural result familiar from thermodynamics where entropy is linked to particle number, as was also shown in \cite{Kandrup:1988zv,Kandrup:1988sg,Prokopec:1992ia}, see \cite{Calzetta:2008iqa} for more discussion and references. The only state without particles is the vacuum state with $|\mathcal{C}(0,0,\cdots)|=1$, which is a pure state with no entropy. This is crucially important for the stability of de Sitter space, which we discuss in the next section. %Strictly speaking, if decoherence is understood as a constraint on the form of the systems density matrix non-zero entropy does not necessarily follow from decoherence, but is an additional assumption for the behaviour.
  
\section{The stability of de Sitter space}
\label{sec:sta}
In the current cosmological picture decoherence of the inflationary perturbations gives rise to the breaking the symmetries of the initial quantum state such that non-trivial time evolution of the system and the formation large scale inhomogeneities is possible \cite{Kiefer:2008ku}, see also the discussion in \cite{Tegmark:2011pi}. The main focus of this work is a very similar effect where again decoherence breaks the symmetries of the state, but here the classical background is exactly de Sitter and the field is conformally coupled with the decohered density matrix as given by (\ref{eq:decQFT0}). In this model any deviation from de Sitter is only visible at the backreaction level, which we study in this section.

Next we calculate the energy-momentum tensor for a conformally coupled scalar field  in the decohered state as described by (\ref{eq:decQFT0}) and study its backreaction in the semi-classical approach. The background spacetime is assumed to be of the flat FLRW form with $ds^2=-dt^2+a(t)^2d\mathbf{x}^2\equiv -dt^2+a^2d\mathbf{x}^2$ and in addition to the quantum field we assume a classical fluid with the equation of state of a cosmological constant, which we call vacuum energy. Even though our interest is in the $4$-dimensional theory, we will write all quantum contributions in $n$ dimensions since that allows us to obtain formally finite results via dimensional continuation, without an explicit violation of general covariance. The various $n$-dimensional quantities used in this section can be found in appendix \ref{sec:A}.

The equation of motion for a conformally coupled scalar field $\hat{\phi}$ is \cite{Birrell:1982ix} 
\ee{\left(-\Box+\xi_c(n) R\right)\hat{\phi}=0\,;\qquad \xi_c(n)\equiv\f{n-2}{4(n-1)}\,, \label{eq:eom}}
where $R$ is the scalar curvature. The solutions can be written as a mode expansion
\ee{\hat{\phi}=\int \f{d^{n-1}\mathbf{k}\, e^{i\mathbf{k\cdot\mathbf{x}}}}{\sqrt{(2\pi)^{n-1}}}\left[\hat{a}_\mathbf{k}^{\phantom{\dagger}}u^{\phantom{\dagger}}_{k}+\hat{a}_{-\mathbf{k}}^\dagger u^{*\phantom{\dagger}}_{{k}}\right]\,\label{eq:adsol2}\, ,}
with the commutation relations $[\hat{a}_{\mathbf{k}}^{\phantom{\dagger}},\hat{a}_{\mathbf{k}'}^\dagger]=\delta^{(n-1)}(\mathbf{k}-\mathbf{k}'),~~[\hat{a}_{\mathbf{k}}^{\phantom{\dagger}},\hat{a}_{\mathbf{k}'}^{\phantom{\dagger}}]=[\hat{a}_{\mathbf{k}}^{{\dagger}},\hat{a}_{\mathbf{k}'}^\dagger]=0$ and where we have defined $k\equiv \vert \mathbf{k}\vert$. Any individual mode solution to (\ref{eq:eom}) can be expressed as 
\ee{u_k=\f{1}{\sqrt{a^{n-1}}}\f{1}{\sqrt{2k/a}}\left(\alpha_k e^{-i \int^t k/a}+\beta_k e^{i \int^t k/a}\right)\, ,\label{eq:bog}}
and can be used to define a vacuum state via
\ee{\hat{a}_{\mathbf{k}}\vert 0 \rangle=0\, .\label{eq:vac}}
Proper normalization of the mode %$ia^{n-1}(\dot{u}^{\phantom{*}}_k u^*_k-u^{\phantom{*}}_k\dot{u}^*_k)=1$, 
requires the coefficients $\alpha_k$ and $\beta_k$ to be related via $\vert\alpha_k \vert^2 -\vert\beta_k \vert^2=1$. The energy-momentum can be written as
\ee{\hat{T}_{\mu\nu}^\phi=-\f{g_{\mu\nu}}{2}\partial_\rho\hat{\phi}\partial^\rho\hat{\phi}+\partial_\mu\hat{\phi}\partial_\nu\hat{\phi} +\xi_c(n)\big[G_{\mu\nu}-\nabla_\mu\nabla_\nu+g_{\mu\nu}\Box\big]\hat{\phi}^2-\delta T_{\mu\nu}\, ,\label{eq:munu}}
where $\delta T_{\mu\nu}$ contains the necessary counter terms \cite{Markkanen:2013nwa}. We can use the semi-classical Einstein equation for calculating the backreaction
%\ee{G_{\mu\nu}M_{\rm pl}^{2} = -\rho_\Lambda g_{\mu\nu}+\langle\hat{T}_{\mu\nu}^\phi\rangle\nonumber\,,} which we can write 
\ea{
\begin{cases}\phantom{-(}3H^2M_{\rm pl}^2&= \langle\hat{T}^\phi_{00}\rangle+\rho_\Lambda\\ -(3H^2+2\dot{H})M_{\rm pl}^2 &= \langle\hat{T}^\phi_{ii}\rangle/a^{2}+p_\Lambda \end{cases}\,,\label{eq:e}}
which leads to the dynamical relation for $H=\dot{a}/a$
\ee{-2\dot{H}M_{\rm pl}^2=\langle\hat{T}^\phi_{00}\rangle+\langle\hat{T}^\phi_{ii}\rangle/a^{2}\,,\label{eq:dyn}}
where we have defined $M_{\rm pl}\equiv (8\pi G)^{-1/2}$. The $\rho_\Lambda=-p_\Lambda$ and $\langle\hat{T}^\phi_{00}\rangle$ are the contributions from the vacuum energy and the conformally coupled scalar field in some yet undefined state. 

Special significance is given to the mode that is made of only the positive frequency contribution as 
\ee{u^{\text{\tiny vac}}_{k}\equiv\f{1}{\sqrt{a^{n-1}}}\f{1}{\sqrt{2k/a}} e^{-i \int^t k/a}%=\lim_{\substack{m\rightarrow0\\ \,\,\,\,\,\,\xi\rightarrow \xi_c(n) }}
\label{eq:bd}\,,}
which defines the so-called conformal vacuum here denoted as $\vert 0^{\text{\tiny vac}}\rangle$, % via $\hat{a}_{\mathbf{k}}^{\text{\tiny vac}}$, 
and which we will simply call the vacuum. The solution (\ref{eq:bd}) has a further special status when $H$ is a constant, since in de Sitter space (\ref{eq:bd}) coincides with the Bunch-Davies (BD) vacuum \cite{Chernikov:1968zm,BD}. In de Sitter space the important characteristics the BD vacuum satisfies are de Sitter invariance and the Hadamard condition \cite{Allen:1985ux}. %and attractor behaviour. 
De Sitter invariance of $\vert 0^{\text{\tiny vac}}\rangle$ leads to the important relation for a constant $H$
\ee{\langle 0^{\text{\tiny vac}}\vert\hat{T}^\phi_{00}\vert 0^{\text{\tiny vac}}\rangle+\langle 0^{\text{\tiny vac}}\vert\hat{T}^\phi_{ii}\vert 0^{\text{\tiny vac}}\rangle /a^2 =0\, ,\label{eq:rp}}
as we will also explicitly verify and which is a consistent solution under backreaction onto the metric as (\ref{eq:dyn}) gives $\dot{H}=0$. This is also true for a massive theory with a non-conformal $\xi$-coupling \cite{markrajan,Anderson:2000wx,Habib:1999cs}. The Hadamard condition essentially states that the short distance behavior of the two-point function should correspond to that in flat space \cite{Bros:1995js}. Imposing that all allowed quantum states have the same ultraviolet (UV) divergences as the BD vacuum state then leads to important equilibration behaviour, where the system always evolves towards the BD vacuum state and hence $H$ always approaches a constant \cite{markrajan,Anderson:2000wx,Habib:1999cs}. As de Sitter space expands exponentially, any change in $H$ induced by a non-de Sitter invariant initial condition is exponentially suppressed on a time scale $\sim 1/H$ \cite{markrajan}. 

The above arguments indicate that the flat exponentially expanding patch of de Sitter space with a non-interacting conformal scalar field is a stable solution in the semi-classical approximation and will persist forever. The process of equilibration into the BD vacuum state could be interpreted as thermalization, but with an important difference: the BD vacuum is expressable as a Fock state with $n_\mathbf{k}=0$ for all $\mathbf{k}$ i.e. it is a pure state with no entropy. For a closed quantum system this is of little relevance, but as we will show, it is crucially important when an environmental sector is included in the discussion.
 
Suppose then that our system is initialized to the vacuum state defined by (\ref{eq:bd}), but after some time instant becomes decohered. In our analysis there will be no need to specify any particular environment, however we emphasize that from the point of view of decoherence theory in order to have a quantum-to-classical transition some form of an environment must exist. We will further assume that there has been sufficient time for the system to decohere such that to a good approximation we may use a diagonal density matrix formed from the eigenstates of particle number as in (\ref{eq:decQFT0}).

Since by definition we only have observable access to the decohered system, we can calculate expectation values simply by using $\hat{\rho}^\text{\tiny D}$
\ee{\langle\hat{\mathcal{O}}\rangle^\text{\tiny D}\equiv{\rm Tr}\big\{\hat{\mathcal{O}}\hat{\rho}^\text{\tiny D}\big\}\,.}
First however we must address what specifically is meant by the number basis used in the expansion for (\ref{eq:decQFT0}). As mentioned, any conformal mode can be expressed as in (\ref{eq:bog}), which essentially is a Bogoluybov transformation of the vacuum.
%\ee{u_k=\f{1}{\sqrt{a^{n-1}}}\f{1}{\sqrt{2k/a}}\left(\alpha_k e^{-i \int^t k/a}+\beta_k e^{i \int^t k/a}\right)\, .\label{eq:bog}}
So for an arbitrary Fock basis used to parametrize $\hat{\rho}^{\text{\tiny D}}$, in (\ref{eq:bog}) we can always define a set of ladder operators such that (\ref{eq:vac}) is satisfied by choosing the mode or rather the $\alpha_k$ and $\beta_k$ accordingly. Effectively, this means that our lack of knowledge of the number basis is only visible as the unknown $\vert\beta_k\vert^2$ coefficient. It is important to note that there is an explicit time dependence in $\hat{\rho}^\text{\tiny D}$. This is because in general the $\mathcal{C}$'s in (\ref{eq:decQFT0}) are time dependent. Additionally, when defining a particle it may be required to transform to a different number basis at each instant in time, a technique which has been formalized and implemented to great sophistication in \cite{Kluger,Habib:1999cs,Dabrowski:2014ica,Anderson:2013ila,Anderson:2013ila1,Dabrowski:2016tsx}, which means that in general also $\alpha_k$ and $\beta_k$ are time-dependent. 

When the $\hat{a}^{\phantom{\dagger}}_\mathbf{k}$ and  $\hat{a}_\mathbf{k}^\dagger$ are chosen to match the number basis of $\hat{\rho}^{\text{\tiny D}}$, we can straightforwardly derive the expectation values in the decohered state
\ea{\langle \hat{a}_\mathbf{q}\hat{a}_\mathbf{k}\rangle^{\text{\tiny D}}&=\langle \hat{a}_\mathbf{q}^\dagger\hat{a}_\mathbf{k}^\dagger\rangle^{\text{\tiny D}}=0\label{eg:a1} \\\langle \hat{a}_\mathbf{q}^\dagger\hat{a}^{\phantom{\dagger}}_\mathbf{k}\rangle^{\text{\tiny D}}&=\delta^{(3)}(\mathbf{k}-\mathbf{q})\langle \hat{a}_\mathbf{k}^\dagger\hat{a}^{\phantom{\dagger}}_\mathbf{k}\rangle^{\text{\tiny D}}\equiv\delta^{(3)}(\mathbf{k}-\mathbf{q})\langle \hat{n}_\mathbf{k}\rangle^{\text{\tiny D}}  \geq 0\label{eg:a2}
%{\rm Tr} \big\{\hat{\rho}^{\text{\tiny D}}\hat{a}_\mathbf{q}^\dagger\hat{a}_\mathbf{k}\big\}={\rm Tr} \big\{\hat{\rho}^{\text{\tiny D}}_\mathbf{k}\hat{a}_\mathbf{q}^\dagger\hat{a}_\mathbf{k}\big\}=\delta^{(3)}(\mathbf{k}-\mathbf{q})\sum_{n_\mathbf{k}=0}^\infty n_\mathbf{k}\vert c(n_{\mathbf{k}},\mathbf{k})\vert^2
\,,}
which follow easily from (\ref{eq:decQFT0}). Using (\ref{eq:munu}), the formulae from Appendix \ref{sec:A} and (\ref{eg:a1}) we get for the energy-density
\ea{\langle\hat{T}_{00}^\phi\rangle^{\text{\tiny D}} &= \f{1}{2}\int\f{d^{n-1}\mathbf{k}}{\sqrt{(2\pi)^{n-1}}}\int\f{d^{n-1}\mathbf{q}}{\sqrt{(2\pi)^{n-1}}}e^{i\mathbf{x}\cdot(\mathbf{k}-\mathbf{q})}\nonumber \\ &\times\bigg\{\dot{u}^{\phantom{s}}_k\dot{u}_q^*+\bigg[\f{\mathbf{k}\cdot\mathbf{q}}{a^2}+2\xi_c(n)(n-1)H\bigg(H\f{n-2}{2}+\f{\partial}{\partial t}\bigg)\bigg]{u}^{\phantom{s}}_k{u}_q^*\bigg\}\nonumber \\&\times\big[\delta^{(3)}(\mathbf{k}-\mathbf{q})+2\langle \hat{a}_\mathbf{q}^\dagger\hat{a}^{\phantom{\dagger}}_\mathbf{k}\rangle^{\text{\tiny D}}\big]-\delta T_{00}\, ,\label{eq:ene}}
which after using (\ref{eq:bog}) and (\ref{eg:a2}) simplifies to
\ee{\langle\hat{T}_{00}^\phi \rangle^{\text{\tiny D}} =\f{1}{2}\int\f{d^{n-1}\mathbf{k}}{(2\pi a)^{n-1}} \f{k}{a}\left(1 +2\vert\beta_k \vert^2\right)\big(1+2\langle \hat{n}_\mathbf{k}\rangle^{\text{\tiny D}} \big)-\delta T_{00}\label{eq:ene1}}
In a similar fashion we can write for the pressure density 
\ea{\langle\hat{T}_{ii}^\phi \rangle^{\text{\tiny D}}/a^2 &= \f{1}{2}\int\f{d^{n-1}\mathbf{k}}{\sqrt{(2\pi)^{n-1}}}\bigg\{\vert\dot{u}_k\vert^2 +\bigg[-\bigg(\f{k}{a}\bigg)^2\f{n-3}{n-1}\nonumber \\&+2\xi_c(n)(2-n)\bigg(H^2\f{n-1}{2}+\dot{H}+H\f{\partial}{\partial t}-(2-n)^{-1}\f{\partial^2}{\partial t^2}\bigg)\bigg]\vert{u}_k\vert^2\bigg\}\nonumber \\&\times\big[1+2\langle \hat{n}_\mathbf{k}\rangle^{\text{\tiny D}}\big]-\delta T_{ii}/a^2\,,}
which also simplifies
\ee{\langle\hat{T}_{ii}^\phi \rangle^{\text{\tiny D}}/a^2 =\f{1}{2}\int\f{d^{n-1}\mathbf{k}}{(2\pi a)^{n-1}} \f{k}{(n-1)a}\left(1 +2\vert\beta_k \vert^2\right)\big(1+2\langle \hat{n}_\mathbf{k}\rangle^{\text{\tiny D}} \big)-\delta T_{ii}/a^2\,.\label{eq:pres}}
%With a very similar derivation one may show that in the BD vacuum one simply has
%\ee{\langle 0^{\text{\tiny BD}}\vert\hat{T}^\phi_{00}\vert 0^{\text{\tiny BD}}\rangle=\f{1}{2}\int\f{d^{n-1}\mathbf{k}}{(2\pi a)^{n-1}} \f{k}{a}\, ;\qquad \langle 0^{\text{\tiny BD}}\vert\hat{T}^\phi_{ii}\vert 0^{\text{\tiny BD}}\rangle/a^{2}=\f{1}{2}\int\f{d^{n-1}\mathbf{k}}{(2\pi a)^{n-1}} \f{k}{(n-1)a}\,.\label{eq:bde}}
Setting $\beta_k=0$ and $\langle \hat{n}_\mathbf{k}\rangle^{\text{\tiny D}}=0$ in (\ref{eq:ene1}) and (\ref{eq:pres}) one obtains the results in the vacuum, %and by a change of integration variables one may recognize that the integrals are identical to what one would obtain in Minkowski space as the zero-point energy and pressure densities of a massless non-interacting particle\footnote{Depending on what is pub...}. With the familiar $n$-dimensional integration formulae (\ref{eq:rp}) is then easily established. Noting this, 
which allows us to write the final form for the energy-momentum tensor
\ee{\begin{dcases}%\langle\hat{T}^\phi_{00}\rangle^{\text{\tiny D}}\equiv
\langle\hat{T}^\phi_{00}\rangle^{\text{\tiny D}}
&=\langle 0^{\text{\tiny vac}}\vert\hat{T}^\phi_{00}\vert 0^{\text{\tiny vac}}\rangle+\int\f{d^{3}\mathbf{k}}{(2\pi a)^{3}} \f{k}{a}N(k,t)\\ \langle\hat{T}^\phi_{ii}\rangle^{\text{\tiny D}}/a^{2} &=\langle 0^{\text{\tiny vac}}\vert\hat{T}^\phi_{ii}\vert 0^{\text{\tiny vac}}\rangle/a^{2}+\int\f{d^{3}\mathbf{k}}{(2\pi a)^{3}} \f{k}{3a}N(k,t)\end{dcases}\,,\label{eq:dece}}
with
\ea{\langle 0^{\text{\tiny vac}}\vert\hat{T}^\phi_{00}\vert 0^{\text{\tiny vac}}\rangle&=\int\f{d^{n-1}\mathbf{k}}{(2\pi a)^{n-1}} \f{k}{2a}-\delta T_{00}\label{eq:vac1}\\ \langle 0^{\text{\tiny vac}}\vert\hat{T}^\phi_{ii}\vert 0^{\text{\tiny vac}}\rangle/a^2&=\int\f{d^{n-1}\mathbf{k}}{(2\pi a)^{n-1}} \f{k}{2(n-1)a}-\delta T_{ii}/a^2\label{eq:vac2}}
and where we have defined
\ee{N(k,t)\equiv \langle \hat{n}_\mathbf{k}\rangle^{\text{\tiny D}}+\vert\beta_k \vert^2\big(1+2\langle \hat{n}_\mathbf{k}\rangle^{\text{\tiny D}} \big)\,,\label{eq:n}}
which can be interpreted as the effective particle number and in terms of $\vert\beta_k\vert^2$ and $\langle \hat{n}_\mathbf{k}\rangle^{\text{\tiny D}}$ is the usual expression encountered in cosmological particle creation, see for example \cite{KeskiVakkuri:1993vk}.
In (\ref{eq:dece}) we have set the dimensions to $n=4$ in the explicit integrals, since they must be convergent. This is comes from assuming that also the decohered state is an UV allowed quantum state for which there cannot be any additional divergences beyond the ones in the vacuum subtracted by $\delta T_{\mu\nu}$ in (\ref{eq:vac1}) and (\ref{eq:vac2}).

We have now shown with respect to a general number basis that when interference involving different particle number eigenstates vanishes, one always obtains an energy--momentum with the expression (\ref{eq:dece}). As mentioned the "00" and "$ii$", or respectively the energy and pressure terms in (\ref{eq:dece}) can be recognized to contain massless particles with the particle number $N(k,t)$.  Since from (\ref{eq:n}) we have $N(k,t)\geq0$, %since the integrals are finite, they must also be greater or equal to zero. T
the special case $N(k,t)=0$ can be  seen via (\ref{eq:decQFT0}) and (\ref{eq:bog}) to correspond to the vacuum i.e the system is pure with strictly zero entropy. When the system is not closed, a natural assumption is that there is at least some entropy %, at the very least due to decoherence and a quantum-to-classical transition, due to quantum entanglement with the environment 
and (\ref{eq:decQFT0}) tells us that for non-zero entropy $N(k,t)$ cannot vanish for all $k>0$. %This leads to the important conclusion already discussed in the previous section that non-zero entropy manifests itself as particle creation. From a statistical point of view this can be understood as the existence of a non-vanishing, however small, probability of observing states other than the vacuum.

By using the standard formulae of $n$-dimensional integration \cite{Peskin:1995ev} one can show that the flat space zero-point energy and pressure of a massive particle sum to zero
\ee{\int \f{d^{n-1 }\mathbf{k}}{(2\pi)^{n-1}}\f{\sqrt{\mathbf{k}^2+m^2}}{2}+\int \f{d^{n-1 }\mathbf{k}}{(2\pi)^{n-1}}\f{\mathbf{k}^2}{2(n-1)\sqrt{\mathbf{k}^2+m^2}}=0\,,\label{eq:rpm}}
and if the counter terms are formed only from covariant tensors introduced by a local action, in de Sitter space one has\footnote{For explicit formulae, see \cite{Markkanen:2013nwa}} $\delta T_{00}+\delta T_{ii}/a^2=0$, which in combination with (\ref{eq:rpm}) allows us to establish (\ref{eq:rp}) from (\ref{eq:vac1}) and (\ref{eq:vac2}). Then finally, by using (\ref{eq:rp}) and (\ref{eq:dece}) we can write from (\ref{eq:dyn}) that if the entropy of the decohered system is non-zero in de Sitter space
\ea{S=-{\rm Tr}\big\{\hat{\rho}^{\text{\tiny D}}\log\hat{\rho}^{\text{\tiny D}}\big\}>0\,,}
there must exist a non-zero particle density for which
\ea{\Rightarrow\quad -2\dot{H}M_{\rm pl}^2= \langle\hat{T}^\phi_{00}\rangle^{\text{\tiny D}}+\langle\hat{T}^\phi_{ii}\rangle^{\text{\tiny D}}/a^{2}&=\f{4}{3}\int\f{d^{3}\mathbf{k}}{(2\pi a)^{3}} \f{k}{a}N(k,t)>0\label{eq:rp2}\\ \Rightarrow\quad \dot{H}&<0\,.\label{eq:rp3}}
%Thus as the main conclusion of this section we have shown that if due to interactions with an environment the system classicalizes to a mixed statistical ensemble of states with definite particle numbers the Hubble rate cannot remain constant. 
Physically, $\dot{H}<0$ follows from the simple fact that for a cosmological fluid made of particles the sum of the energy and pressure never cancels. The fact that a particle density in de Sitter space leads to a time evolution of the Hubble rate is well-known and for example has recently been studied in \cite{clif}, but the connection to information loss and decoherence is new and our main result. Equation (\ref{eq:rp3}) signals an instability with respect to backreaction and as we will elaborate more in the next section, in an expanding space the particle production resulting in a non-zero $N(k,t)$ is continuous. Over long periods of time this can lead to a significant reduction in $H$, which for a Universe dominated by dark energy with $\rho_\Lambda=-p_\Lambda$ implies a distinctively different fate to a classical approximation.

Particle creation and the decrease of the Hubble rate in de Sitter space in many ways resembles the decrease of an electric field due to backreaction of particles that result from Schwinger pair creation \cite{Anderson:2013ila1,Anderson:2013ila}. For a semiclassical backreaction analysis of the Schwinger effect, see \cite{Kluger:1992gb}.

In order to derive (\ref{eq:rp3}) we did not need explicit expressions for the counter terms due to their cancellation in de Sitter in (\ref{eq:rp2}). This is only true for a constant $H$ and for studying the implications of the instability, we must generalize our results for a dynamical Hubble rate. Fortunately out of all the relations we used, $\delta T_{00}+\delta T_{ii}/a^2=0$ is the only one that is contingent on the constancy of $H$ and in particular the results (\ref{eq:dece}) hold for an arbitrary scale factor $a$. So any potential difference to de Sitter space comes from the counter terms and is convergent for the already mentioned reason that for UV allowed states the divergences, as given by the integrals in (\ref{eq:vac1}) and (\ref{eq:vac2}), must be the same.

For deriving  expressions for the $\delta T_{\mu\nu}$, perhaps the most convenient way is the adiabatic subtraction method \cite{Parker:1974qw,Parker:1974qw1,Bunch:1980vc}. As this is a standard technique in curved space quantum field theory, here we will omit the explicit derivation of the counter terms and refer the reader to \cite{Parker:2009uva,Birrell:1982ix} for details. 

One may show by straightforward but slightly tedious calculations that for the conformally coupled case adiabatic subtraction results in the counter terms 
\ea{\delta T_{00}&=%\lim_{m\rightarrow 0}\f{1}{2}\int\f{d^{3}\mathbf{k}}{(2\pi a)^{3}}\bigg\{\omega +\frac{m^4 \dot{a}^2}{8 \omega ^5a^2}+\frac{m^4 \ddot{a}^2}{32 \omega^7a^2 }-\frac{105 m^8 \dot{a}^4}{128  \omega^{11}a^4}+\frac{7 m^6 \dot{a}^4}{8 \omega^9a^4 }-\frac{m^4 \dot{a}^4}{8 \omega ^7a^4}- \frac{m^4 a^{(3)}\dot{a}}{16  \omega^7a^2}\nonumber \\&+\frac{7 m^6 \dot{a}^2 \ddot{a}}{16 \omega^9a^3}-\frac{5 m^4 \dot{a}^2 \ddot{a}}{16 \omega^7a^3}\bigg\}=
\int\f{d^{n-1}\mathbf{k}}{(2\pi a)^{n-1}} \f{k}{2a}+\f{1}{480\pi^2}\bigg[\frac{\ddot{a}^2}{2 a^2}+\frac{\dot{a}^4}{a^4}-\frac{a^{(3)} \dot{a}}{a^2}-\frac{\dot{a}^2 \ddot{a}}{a^3}\bigg]\,,\label{eq:conf1}\\\delta T_{ii}/a^2%&=\lim_{m\rightarrow 0}\f{1}{2}\int\f{d^{3}\mathbf{k}}{(2\pi a)^{3}}\bigg\{\frac{\omega}{3}-\frac{m^2}{3 \omega }-\frac{m^4 \ddot{a}}{12  \omega ^5a}+\frac{5 m^6 \dot{a}^2}{24 \omega^7 a^2}-\frac{m^4 \dot{a}^2}{8  \omega^5a^2}+\frac{m^4 a^{(4)}}{48 \omega ^7a}-\frac{7 m^6 \ddot{a}^2}{32  \omega^9a^2}\nonumber \\&+\frac{5 m^4 \ddot{a}^2}{32  \omega^7a^2}-\frac{385 m^{10} \dot{a}^4}{128  \omega^{13}a^4}+\frac{581m^8 \dot{a}^4}{128  \omega^{11}a^4}-\frac{7 m^6 \dot{a}^4}{4  \omega^9a^4}+\frac{m^4 \dot{a}^4}{8 \omega^7a^4 }-\frac{7 m^6 a^{(3)}(t) \dot{a}}{24 \omega^9a^2 }+\frac{5 m^4 a^{(3)} \dot{a}}{24  \omega^7a^2}\nonumber \\&+\frac{77 m^8 \dot{a}^2 \ddot{a}}{32 \omega^{11} a^3}-\frac{133 m^6 \dot{a}^2 \ddot{a}}{48  \omega^9a^3}+\frac{7 m^4 \dot{a}^2 \ddot{a}}{12  \omega^7a^3}\bigg\}\nonumber\\&
&=\int\f{d^{n-1}\mathbf{k}}{(2\pi a)^{n-1}} \f{k}{2(n-1)a}+\f{1}{1440\pi^2}\bigg[\frac{a^{(4)}}{a}+\frac{3 \ddot{a}^2}{2 a^2}+\frac{\dot{a}^4}{a^4}+\frac{2 a^{(3)} \dot{a}}{a^2}-\frac{4 \dot{a}^2 \ddot{a}}{a^3}\bigg]\,,\label{eq:conf2}}
with which we can generalize the right-hand side of (\ref{eq:rp2}) as
\ea{-2\dot{H}M_{\rm pl}^2=%\langle\hat{T}^\phi_{00}\rangle^{\text{\tiny D}}+\langle\hat{T}^\phi_{ii}\rangle^{\text{\tiny D}}/a^{2}=
\f{4}{3}\int\f{d^{3}\mathbf{k}}{(2\pi a)^{3}} \f{k}{a}N(k,t)-\f{1}{480\pi^2}\bigg[\frac{2}{3} H^2 \dot{H}+2 \dot{H}^2+H \ddot{H}+\frac{1}{3} H^{(3)}\bigg]\,.\label{eq:rp4}}
The additional finite contribution from the counter terms in the square brackets of (\ref{eq:rp4}) is a manifestation of the conformal anomaly\footnote{Using (\ref{eq:conf1}) and  (\ref{eq:conf2}) in (\ref{eq:vac1}) and (\ref{eq:vac2}) we can verify the result for conformally flat spaces \cite{Birrell:1982ix} \ee{g^{\mu\nu}\langle 0^{\text{\tiny vac}}\vert\hat{T}^\phi_{\mu\nu}\vert 0^{\text{\tiny vac}}\rangle=\f{1}{480\pi^2}\bigg[\frac{3 \dot{a}^2 \ddot{a}}{a^3}-\frac{a^{(4)}}{a}-\frac{\ddot{a}^2}{a^2}-\frac{3 a^{(3)} \dot{a}}{a^2}\bigg]=\f{1}{2880\pi^2}\bigg[G_{\mu\nu}G^{\mu\nu}-\f{R^2}{3}+\Box R\bigg]\nonumber}
}, which is discussed in \cite{Capper:1974ic,Capper:1975ig,Deser:1976yx} and studied via the adiabatic approach in \cite{Bunch:1978gb,Bunch:1980vc,Markkanen:2013nwa}.

We conclude this discussion on a small technical note concerning our use of analytic continuation to $n$-dimensions. As is apparent from the integrals in (\ref{eq:rpm}), the cancellation of the vacuum energy and pressure density components is not true in the cut-off sense, which has sometimes been argued to have relevance for the stability of de Sitter space, see \cite {Maggiore:2010wr2} for references and counterarguments. But as we may see from the adiabatic counter terms (\ref{eq:conf1}) and  (\ref{eq:conf2}), also the combination $\delta T_{00}+\delta T_{ii}/a^2$ does not vanish for a cut-off in de Sitter and this violation precisely cancels that of the vacuum contributions such that (\ref{eq:vac1}) and (\ref{eq:vac2}) will nonetheless lead to the cancellation (\ref{eq:rp}). %To summarize, the usual approach of adiabatic subtraction with formally divergent 4-dimensional integrals would also also lead to equation (\ref{eq:rp4}), even though separately the vacuum contribution and the counter term contribution are not de Sitter invariant.

We can now proceed to a self-consistent backreaction analysis.

\section{Implications for the fate of the Universe} 
\label{sec:ti}
In this section we give an estimate for the evolution of the Hubble rate as implied by the analyses of sections \ref{sec:dec} and \ref{sec:sta}.

In order to perform the calculation in closed from, we first need to simplify the expression (\ref{eq:rp4}). A natural first approximation is to study solutions where on the right-hand side of (\ref{eq:rp4}) we assume $H$ to be to a good approximation constant. This comes from the expectation that backreaction is a weak effect, which only after a long time can result in a significant deviation from de Sitter.

As we have already mentioned, given our assumptions of the UV behaviour of the theory in a closed system a quantum field in de Sitter space tends to equilibrate
towards the BD vacuum. This is evident from (\ref{eq:rp2}): any initial condition with a non-zero particle density, $N(k,t_0)=N(k)$, will decay as $1/a^4$ and quickly lead to $\dot{H}\sim0$. As the BD vacuum is a pure state, particle creation must be continuous in order counteract this effect and maintain non-zero entropy in the system. Note that this is ultimately a local statement for a local observer who can probe only a finite volume in space. More concretely if for completeness we take the accessible volume to be the Hubble sphere $\propto H^{-3}$, if new particles are not continuously created the number of particles and thus entropy inside this volume will be driven towards zero by the expansion of space. But  if there is any decoherence or loss of information as discussed in section \ref{sec:dec}, the entropy cannot be zero. In a way, the equilibration of the quantum field and information loss are two opposite and competing forces.

If we assume that particles are produced at a constant rate we can conclude that the energy and pressure densities then approach, possibly $H$ dependent, constants. One can easily understand this by solving the late time limit of the evolution for radiation in the presence of a constant source term $\nu$ 
\ee{\partial_t\langle{\hat{T}}^\phi_{00}\rangle^{\text{\tiny D}}=-4H\langle{\hat{T}}^\phi_{00}\rangle^{\text{\tiny D}}+\nu\,\quad \Rightarrow\quad \langle{\hat{T}}^\phi_{00}\rangle^{\text{\tiny D}}\longrightarrow\f{\nu}{4H}.\label{eq:eveq}}
If there is no $H$ dependence in $\langle{\hat{T}}^\phi_{00}\rangle^{\text{\tiny D}}$  and $\langle{\hat{T}}^\phi_{ii}\rangle^{\text{\tiny D}}/a^2$ the right-hand side of the backreaction equation (\ref{eq:rp4}) implies an unbounded monotonically decreasing solution, unless the higher order derivative terms in the square brackets of (\ref{eq:rp4}) become dominant and stabilize the solution once the evolution is no longer to a good approximation de Sitter. Both scenarios seem rather unphysical, the first simply because it is unstable and the latter because it would mean that the fate of the Universe is crucially dependent on the contribution from the conformal anomaly. At any rate, both would imply that eventually the backreaction will drive the Universe away from the de Sitter geometry.
 
A more natural set of solutions is found when the saturated limits of the energy and pressure densities are assumed to be functions of $H$ that vanish at $H\rightarrow0$ since then the evolution of $H$ is necessarily well-behaved and particle creation stops only when the curvature of space vanishes. We will first focus on the case where the only scale in the energy and pressure densities is $H$, i.e. one sets $N(k,t)=N(k/(aH))$, which leads to $\langle{\hat{T}}^\phi_{00}\rangle^{\text{\tiny D}}\sim\langle{\hat{T}}^\phi_{ii}\rangle^{\text{\tiny D}}/a^2\sim H^4$. Neglecting the derivatives of $H$ on the right-hand side of (\ref{eq:rp4}) allows us then to write
\ee{-2\dot{H}M_{\rm pl}^2=\f{2b}{3} H^4\quad\Rightarrow\quad H=\f{H_0}{\Big(b\f{H_0^3}{M_{\rm pl}^2}t+1 \Big)^{1/3}}\,,\label{eq:Hev}}
where $H_0\equiv H(0)$ and $b$ is some arbitrary positive number. As an important special case consistent with all our assumptions we can use a thermal density matrix for a massless field with the Gibbons-Hawking de Sitter temperature $T_{\rm dS}=H/(2\pi)$ \cite{Gibbons:1977mu} for calculating the value of the $b$ parameter: 
\ee{N(k,t)=\left(e^{\f{2\pi k}{aH}}-1\right)^{-1}\quad\Rightarrow\quad b=1/(240\pi^2)\,.\label{eq:GH}}
We note that based on an analogy with black hole evaporation, the result (\ref{eq:Hev}) with (\ref{eq:GH}) has already been obtained in section 6.2 of \cite{Padmanabhan:2003gd} (see also the discussion in \cite{Padmanabhan:2002ha,Rigopoulos:2013exa}) and studied in the inflationary context in \cite{clif}. Furthermore, recently in \cite{Markkanen:2016vrp} this result was derived from first principles by tracing over the degrees of freedom beyond the horizon.

For the late time Universe with a small $H$, which is our primary interest, we see that since $-\dot{H}/H^2\sim H^2/M_{\rm pl}^2$ from (\ref{eq:Hev}) our assumption of neglecting the derivatives of $H$ is in fact a reasonable approximation for all times as for the current Hubble rate one has $H_0\sim 10^{-42}$GeV and since $H\leq H_0$. We can also verify this by using the $H$ obtained from (\ref{eq:Hev}) in the terms in the square brackets of (\ref{eq:rp4}) 
\ee{{H^{-4}}\bigg[\frac{2}{3} H^2 \dot{H}+2 \dot{H}^2+\frac{1}{3} H^{(3)}+H \ddot{H}\bigg]%=-\frac{2b }{9}\bigg(\f{H}{M_{\rm pl}}\bigg)^2+\cdots
%+\mathcal{O}\bigg(\f{H_0^4}{M_{\rm pl}^4}\bigg)
\sim\bigg(\f{H}{M_{\rm pl}}\bigg)^2\,.\label{eq:smalls}}
It is possible to iteratively solve equation (\ref{eq:rp4}) to include contributions from derivatives of $H$ as an expansion around (\ref{eq:Hev}) as shown in Appendix \ref{sec:B}, however this is not in practice needed for a small $H$.

From (\ref{eq:Hev}) it is trivial to find the behaviour of the scale factor to be
\ee{a(t)=\exp\bigg\{\f{3 M_{\rm pl}^2}{2bH_0^2}\bigg[\bigg(\f{bH_0^3 }{M_{\rm pl}^2}t+1\bigg)^{2/3}-1\bigg]\bigg\}\,,\label{eq:aa}}
which we have normalized as $a(0)=1$. What the solutions (\ref{eq:Hev}) and (\ref{eq:aa}) imply is that the expansion of the Universe will continue forever, but in contrast to the purely classical approximation, now $H$  asymptotically approaches 0. The dynamics of the expansion can be seen to have two distinct phases, starting with a constant $H\sim H_0$ and ending with $H\sim (M_{\rm pl}^2/t)^{1/3}$. This means that again there exists a time after which the geometry can no longer be viewed as de Sitter. We can quantify this time scale by calculating the half-life 
\ee{H(t_{1/2})=H_0/2\quad\Leftrightarrow\quad t_{1/2}=\f{7}{b}\f{M_{\rm pl}^2}{H_0^2}H_0^{-1}\,.\label{eq:hl}}
For the current value for the Hubble rate %$H_0\sim 10^{-42}$ GeV
we get an estimate for $t_{1/2}$ that is more than $10^{100}$-times the age of the Universe, for $b\sim\mathcal{O}(1)$. From the point of view of feasible experiments this implies that for the late time Universe this effect is unobservable, at least for the solutions with $N(k,t)\approx N(k/(aH))$ in (\ref{eq:rp4}).

An obvious question we must now address is the one regarding the source of the continuous particle creation. %Since by definition we can only observe the quantum field in the decohered state as all other degrees of freedom are coarse grained over, 
At least in the effective sense we can resolve this issue by invoking consistency: since $\nabla^\mu G_{\mu\nu}=0$ by construction for Einsteinian gravity, covariant conservation must be respected by the energy-momentum, which then gives
\ee{\nabla^\mu(-g_{\mu\nu}\rho_\Lambda+\langle\hat{T}_{\mu\nu}^\phi\rangle^\text{\tiny D})=0\quad\Leftrightarrow\quad \dot{\rho}_\Lambda=-3H\left(\langle\hat{T}_{00}^\phi\rangle^\text{\tiny D}+\langle \hat{T}_{ii}^\phi\rangle^\text{\tiny D}/a^2\right)=-2bH^5\,,\label{eq:decl}}
where we have used $\langle\hat{T}_{00}^\phi\rangle^\text{\tiny D}\sim H^4$ and neglected the sub-leading $\mathcal{O}(\dot{H})$ term. What then follows is that the only way to satisfy covariant conservation for $\rho_\Lambda=-p_\Lambda$ is to have $\dot{\rho}_\Lambda<0$ and physically we can interpret the vacuum energy as a gradually emptying reservoir providing the energy required for the creation of the particles. By using (\ref{eq:eveq}) in (\ref{eq:decl}) we can also deduce that the change in $\rho_\Lambda$ can be seen as the particle creation flux, $-\dot{\rho}_\Lambda=\nu$. %We stress that this interpretation should not be taken literally and it is crucially dependent on covariant conservation.

As discussed in \cite{Spradlin:2001pw} the possibility of a fundamental time evolution in $\rho_\Lambda$ seems a priori problematic as classically the vacuum energy is simply a parameter of the Lagrangian, the cosmological constant. It is however important to realize that this effect is only seen in the quantised theory, for which there is no longer necessarily a direct equality between a constant parameter of the Lagrangian and a physically measurable contribution in the semi-classical Einstein equation. Because of this, one may argue $\rho_\Lambda$ and $p_\Lambda$ must be viewed as effective quantities that in principle can have time dependence. A good example of a similar discrepancy is the running of constants due to quantum corrections, which is well-known in the context of particle physics and for an interacting theory on a curved background running also occurs for the gravitational parameters including the cosmological constant, see for example \cite{Markkanen:2014poa,Sola:2013gha,Elizalde:1993ew,Elizalde:1993qh,Shapiro:2000dz}. 

In a complete first principle analysis however, one would expect that covariant conservation is satisfied without the need to assign any time dependence in $\rho_\Lambda$. By allowing more complicated forms for $N(k,t)$ one may show that a time evolution in $\rho_\Lambda$ is not in general required for covariant conservation. We can show this by writing an ansatz, $N(k,t)\rightarrow\left(1+(2/b)F(t)H^{-4}\right)N(k/(a H))$, where $F(t)$ is a small function with the initial condition $F(0)=0$, $N(k,t)$ is parametrized in terms of $b$ as in (\ref{eq:Hev}) and demanding covariant conservation with a constant $\rho_\Lambda$. Close to de Sitter one may easily find an approximate solution for $F(t)$. Quite naturally, for small times one obtains $F(t)=-2b H^5 t$, which results approximately in the evolution (\ref{eq:Hev}) and also a decrease of the energy-density as $\langle\dot{T}^\phi_{00}\rangle\sim-2b H^5 $, very much like in (\ref{eq:decl}), but with a strictly constant $\rho_\Lambda$. Thus for such an ansatz the dynamics are essentially identical to the case $N(k,t)=N(k/(aH))$, when the system behaves approximately as de Sitter space.

As a final note, we comment on the limit $t\rightarrow \infty$ of our results. Importantly, we have shown here and in more detail in Appendix \ref{sec:B} that for $N(k,t)=N(k/(aH))$ (\ref{eq:rp4}) can be naturally solved via an expansion in terms of a very small quantity ${H_0^2}/M_{\rm pl}^2$, but it is not obvious that $N(k,t)=N(k/(aH))$ is a good approaximation when the system has evolved significantly away from de Sitter, even if this was initially true. Furthermore, we have not properly addressed the convergence of this series. As (\ref{eq:rp4}) does contain a third order derivative term, some solutions might be subject to an Ostrogradsky instability, although it is questionable whether they can be viewed as physical \cite{Parker:1993dk}. %Analysing these delicate issues we leave for future work.

\section{Discussion}
The decoherence program has provided a compelling explanation as to why out of all the states allowed by quantum theory only the ones we may characterize as classical are frequently encountered in Nature. The key realization is that macroscopic systems are never closed but continuously influenced by their environment. This leads to the suppression of quantum interference. As this continuous environmental monitoring alters the state of the system also the gravitational backreaction changes.

In this work we have studied backreaction via the semi-classical Einstein equation for a model consisting of a conformally coupled scalar field and classical vacuum energy in the presence of a decoherence inducing environment. Our approach was to postulate that %particle number is continuously monitored by the environment such that in the density matrix describing the system, 
interference between particle number eigenstates is negligible producing a classical statistical ensemble of states with definite particle number. This was motivated by the fact that at least for the current Universe with weak curvature, to a good approximation particles are robust classical observables. Choosing the pointer basis to be something different than the number basis is likely to lead to different results. For example, the classicalization of the primordial pertubations from inflation is believed to take place in the field amplitude basis \cite{Polarski:1995jg,Kiefer:2006je,Kiefer:2008ku,Burgess:2006jn}, although with a similar violation of de Sitter invariance \cite{Kiefer:2008ku}. Precisely determining a pointer basis for a conformal scalar field in the late time Universe presumably would require a first principle analysis with a specific chosen form for the environmental interaction.

For a closed system with a conformal scalar field on a de Sitter background the Bunch-Davies vacuum acts as an equilibrium state and is a stable solution under semi-classical backreaction \cite{markrajan}. But it is also a state with no entropy and in the presence of an environment that decoheres the system leading to particles as classical observables any entropy generation results in particle creation and a violation of de Sitter invariance. Because of the competing nature of the two effects, the equilibration of the quantum field into a pure state due to the expansion of space and the transforming of pure states into mixed ones via decoherence, the particle creation is expected to be continuous. This provides a mechanism by which over long periods of time the Hubble rate can be significantly reduced, which is the main conclusion of this work. 

By using the semi-classical approach we found a self-consistent solution for the Hubble rate in the presence of constant particle creation that after time scales $\sim10^{100}$-times the age of the Universe modified the dynamics of a spacetime intialized to de Sitter to eventually behave as $H\sim(M_{\rm pl}^2/t)^{1/3}$. We argued via consistency that this has an interpretation as the decay of the vacuum energy and has important implications for the fate of a Universe dominated by dark energy.

Decoherence can be seen to arise from irreversible delocalization of information from the system into the unobservable environment. In this work we left the environment unspecified, but in order for this process to be possible some type of environmental degrees of freedom must exist. For a local observer it seems natural to assume observational access to only a subset of all possible degrees of freedom. For example, spacelike separations of events and, in particular for de Sitter space, particle horizons obviously impose loss of correlations. %We note that an interesting approach for studying the backreaction implications from loss of information due to the de Sitter horizon would be to implement the techniques developed for entanglement entropy in \cite{Maldacena:2012xp}. %Furthermore, as gravity interacts with everything the gravitational degrees of freedom could provide the means via which any quantum field may disperse information. Gravitational decoherence is not a new idea \cite{Karolyhazy:1966zz,Diosi:1988uy,Penrose:1996cv} and is also currently actively studied, see \cite{Blencowe:2012mp} for more references. A closer inspection of this assertion likely requires one to go beyond the semi-classical approach. 
Macroscopic observers probing the particle content of the Universe such as ourselves may as well play a role, in particular for inducing classicalization in the particle number basis.

Although our discussion was kept free of a reference to a thermodynamic equilibrium, due to the deep connection between horizons and thermodynamics \cite{Gibbons:1977mu,Padmanabhan:2002sha,Padmanabhan:2003gd,Padmanabhan:2009vy} it is possible that also here a thermal interpretation is applicable. The "hot tin can" description valid for de Sitter space in static coordinates \cite{kof,Kaloper:2002cs,Susskind:2003kw} when implemented in an expanding space has already been shown to lead to very similar conclusions as presented here \cite{clif}. On general grounds a thermodynamic instability of de Sitter space was argued to exist in \cite{Antoniadis:2006wq,Mottola:1985qt,Sekiwa:2006qj}. %We also point out that the response of a co-moving Unruh-DeWitt particle detector in de Sitter space with a conformal scalar field is equivalent to that of a constant thermally distributed particle density \cite{Birrell:1982ix} (see \cite{Garbrecht:2004du} for a recent discussion).

\acknowledgments{The author would like to thank Malcolm Fairbairn, Eugene Lim and Arttu Rajantie for discussions, and Robert Brandenberger, Claus Kiefer and Thanu Padmanabhan for a critical reading of the manuscript. The research leading to these results has received funding from the European Research Council under the European Union's Horizon 2020 program (ERC Grant Agreement no.648680).}

\appendix
\section{Tensor formulae}
\label{sec:A}
In this appendix for completeness we give the necessary $n$-dimensional quantities used in section \ref{sec:sta}. For a more complete list see \cite{Markkanen:2013nwa}.

In the flat FLRW coordinates with $ds^2=-dt^2+a(t)^2d\mathbf{x}^2$ in the $(+,+,+)$ conventions of \cite{Misner:1974qy} we can write in $n$ dimensions
\ea{\Box f(t)&=-\ddot{f}(t)+(1-n)\f{\da}{a}\dot{f}(t),\\(-\nabla_0\nabla_0+g_{00}\Box)f(t)&=(n-1)\f{\da}{a}\dot{f}(t), \\
(-\nabla_i\nabla_i+g_{ii}\Box)f(t)&=a^2\bigg[(2-n)\f{\da}{a}\dot{f}(t)-\ddot{f}(t)\bigg],\\
R&=2(n-1)\bigg(\f{\da^2}{a^2}+\f{\dda}{a}\bigg)+(n-1)(n-4)\f{\da^2}{a^2}\label{eq:Rt},\\
G_{00}&=\f{(n-1)(n-2)}{2}\bigg(\f{\dot{a}}{a}\bigg)^2,\\
G_{ii}&=a^2(2-n)\bigg[\f{(n-3)}{2}\bigg(\f{\dot{a}}{a}\bigg)^2+\f{\ddot{a}}{a}\bigg],}
where $a(t)\equiv a$.

\section{Higher order corrections}
\label{sec:B}
In order to obtain the next-to-leading correction to (\ref{eq:Hev}) we can write an ansatz
\ee{H=\f{H_0}{\Big(b\f{H_0^3}{M_{\rm pl}^2}t+1 \Big)^{1/3}}+\delta H+\cdots\,,\label{eq:h1}}
where $\delta H$ is now a small perturbation that vanishes at $t=0$.
By treating the contribution in the square brackets of (\ref{eq:rp4}) 
%\ee{-2\dot{H}M_{\rm pl}^2=\f{2b}{3} H^4-\f{1}{480\pi^2}\bigg[\frac{2}{3} H^2 \dot{H}+2 \dot{H}^2+\frac{1}{3} H^{(3)}+H \ddot{H}\bigg]\,,\label{eq:Hev3}} 
and $\delta H$ as small numbers suppressed by $\sim  H_0^2/M_{\rm pl}^2$ allows us to write an equation for $\delta H$ as
\ee{-2M_{\rm pl}^2\delta\dot{H}=\delta{H}\f{8bH_0^3}{3\Big(b\f{H_0^3}{M_{\rm pl}^2}t+1 \Big)}+\f{b H_0^6}{2160\pi^2M_{\rm pl}^2\Big(b\f{H_0^3}{M_{\rm pl}^2}t+1 \Big)^2}+\mathcal{O}\left(H_0^4\f{H_0^4}{M_{\rm pl}^4}\right)\,,}
which for the boundary condition $H(0)=H_0$ has the solution
\ee{\delta H=\Bigg\{\f{H_0}{1440\pi^2\Big(b\f{H_0^3}{M_{\rm pl}^2}t+1 \Big)^ {4/3}}-\f{H_0}{1440\pi^2\Big(b\f{H_0^3}{M_{\rm pl}^2}t+1 \Big)}\Bigg\}\bigg(\f{H_0}{M_{\rm pl}}\bigg)^2\label{eq:h2}\,,} 
with which one may verify that the error in (\ref{eq:rp4})  is indeed reduced from $\mathcal{O}\big({H_0^2}/{M_{\rm pl}^2}\big)$ to $\mathcal{O}\big({H_0^4}/{M_{\rm pl}^4}\big)$. It is also straightforward to iterate (\ref{eq:rp4})  further in order to obtain the higher order corrections. As in (\ref{eq:hl}) of section \ref{sec:ti} we can calculate the half-life $H(t_{1/2})=H_0/2$ from (\ref{eq:h1}), which now gets a small additional contribution from $\delta H$ as
\ee{t_{1/2}=\f{7}{b}\f{M_{\rm pl}^2}{H_0^2}H_0^{-1}\bigg(1-\f{H_0^2}{3360\pi^2M_{\rm pl}^2}\bigg)+\mathcal{O}\bigg(H_0^{-1}\f{H_0^2}{M_{\rm pl}^2}\bigg)\,.}

This procedure of iteratively solving the semi-classical Einstein equation is discussed in \cite{Parker:1993dk} and it has the great advantage that it circumvents the difficult problem of providing initial conditions for the higher order derivative terms introduced by the quantum corrections.

\end{document}